\documentclass[prd,article,a4paper,twocolumn,superscriptaddress]{revtex4}
\usepackage{graphicx}
\begin{document}

\newcommand{\be}{\begin{equation}}
\newcommand{\ee}{\end{equation}}
\newcommand{\bq}{\begin{eqnarray}}
\newcommand{\eq}{\end{eqnarray}}
\newcommand{\bsq}{\begin{subequations}}
\newcommand{\esq}{\end{subequations}}
\newcommand{\bc}{\begin{center}}
\newcommand{\ec}{\end{center}}

\title{Fractal Properties and Small-scale Structure of Cosmic String Networks}
\author{C.J.A.P. Martins}
\email[Electronic address: ]{C.J.A.P.Martins@damtp.cam.ac.uk}
\affiliation{Centro de F\'{\i}sica do Porto, 
Rua do Campo Alegre 687, 4169-007, Porto, Portugal}
\affiliation{Department of Applied Mathematics and Theoretical Physics, Centre for Mathematical Sciences,\\ University of Cambridge, Wilberforce Road, Cambridge CB3 0WA, United Kingdom}
\author{E.P.S. Shellard}
\email[Electronic address: ]{E.P.S.Shellard@damtp.cam.ac.uk}
\affiliation{Department of Applied Mathematics and Theoretical Physics,
Centre for Mathematical Sciences,\\ University of Cambridge,
Wilberforce Road, Cambridge CB3 0WA, United Kingdom}

\date{29 November 2005}
\begin{abstract}
We present results from a detailed numerical study of the small-scale and loop production properties of cosmic string networks, based on the largest and highest resolution 
string simulations to date. We investigate the
non-trivial fractal properties of cosmic strings, in particular, 
the fractal dimension and renormalised string mass per unit
length, and we also study velocity correlations.  We demonstrate important differences between string networks in flat (Minkowski) spacetime and the two very similar expanding cases.  For high resolution matter era network simulations, we provide 
strong evidence that small-scale structure has converged to `scaling' on all dynamical 
lengthscales, without the need for other radiative damping mechanisms.  We also discuss
preliminary evidence that the dominant loop production size is also approaching scaling.  
\end{abstract}
\preprint{DAMTP-2005-???}
\maketitle

\section{\label{sint}Introduction}

Topological defects are an unavoidable consequence of
phase transitions in the early universe (for a review see ref.~\cite{vsh}). 
Cosmic strings, in particular,
will form in a range of cosmological scenarios, including at the end of brane inflation \cite{tye0,tye1,cope}. Since they are intrinsically
nonlinear objects, one is ultimately forced to resort to high resolution numerical
simulations \cite{bb,as} if one wants to study their evolution and cosmological
consequences in detail.

In recent years, while some
authors have exploited ever-increasing computing
capabilities \cite{cam,moore,prl},
many others have tried to simplify the problem. For example, a number of
flat (Minkowski) spacetime string simulations have been
performed \cite{flat,vhs1,vhs2,pst,chm,olum}, which it has been argued
are a good approximation to the expanding case.
(The idea is that the string correlation length is significantly smaller than
the horizon size and so the network should not feel the expansion.) 
Another popular approach is to generate artificial
networks \cite{abr1,pv} made of a (variable) number of segments with a
given size and velocity, and aiming to mimic the evolution of a real
network by enforcing (by hand) suitable
variations of these parameters (e.g., as predicted by quantitative
analytic string evolution models \cite{ack,ms1a,ms1b,extend}).
Finally, doubts have also been raised as to whether  the Nambu action accurately 
reproduces the dynamics of vortex-strings in the underlying field 
theory \cite{vhs1,bevis} (though we will not discuss this issue further here, 
see \cite{moore}).  The problem is that different approaches yield quantitatively
different and even sometimes contradictory cosmological consequences.
This raises the question of how reliable all these approximations really are. 
One of the goals of this letter is to clarify these issues. In particular, we argue that there are important 
differences between string network evolution in flat spacetime and in an 
expanding universe.

There is also considerable confusion in the literature on the issue of scaling. This is partially due to the fact that people define scaling in different ways. The weakest definition -- large-scale scaling -- is simply to say that one wants the energy density to be a constant fraction of the total density in the universe. This is the attractor solution of the evolution of long-string networks, and is strongly confirmed by both analytical and numerical evidence. On the other hand, this by itself does not guarantee that small scale features of the network are also scaling, such as the typical scale of small scale wiggles, that of loop production, or even the correlation length itself. In this letter we aim to clarify these different concepts. We also present strong evidence for scaling of the small-scale features of the string network (as measured by its multifractal dimension \cite{frac1,frac2,frac3} and renormalized mass per unit length) and we discuss the scaling of the characteristic loop production size.

\section{\label{sfra} Averaged and microscopic properties}

We have performed many thousands of CPU hours of
cosmic string simulations, using an upgraded version of the Allen-Shellard
string code on the COSMOS supercomputer. Some preliminary, large-scale
results have already been discussed in \cite{moore,prl}, and a thorough
description of the simulations will be presented elsewhere \cite{prep2}.
Here, however, we take advantage of the extremely high resolution of our
runs to provide a detailed description of the
averaged and small-scale properties of cosmic string networks, as well as of the processes that determine loop production.  This goes substantially further than the previous analyses of 
refs.~\cite{asconf,bbconf} and it is interesting to compare with the more recent flat 
spacetime analysis of ref.~\cite{olum}. 
Ultra-high resolution simulations were performed in the matter
and radiation epochs
and in flat (Minkowski) spacetime. The initial Vachaspati-Vilenkin
networks \cite{vv} have resolutions between 75 and 256 points per correlation length (PPCL), and we subsequently enforce a constant resolution in physical coordinates. This resolution is a substantial improvement over currently published simulations. These networks are then evolved for a dynamic range (in conformal time) of order $3$ (and up to $6$), each taking 
several weeks of processor time. 

We note that not everybody agrees on when a network
reaches the scaling regime.
For example, numerically we find that the number of strings per
horizon $\rho t^2/\mu_0$ becomes a constant (a
plausible definition of linear scaling)
before the correlation length $\xi$ becomes a
constant fraction of physical time $t$ (which is another possibility).
This, in turn, happens before other properties of the network reach
scaling---indeed, there are
a number of different timescales involved in the approach to scaling.
We have developed our own statistics, based entirely on \textit{small-scale}
properties, to quantify how far a network is from scaling.
We will describe them in detail elsewhere \cite{prep2}, but for the
purposes of this letter it is enough to think of the
two definitions above as roughly measuring \textit{large-scale} and
\textit{small-scale} scaling. Our high resolution and
large dynamic range allow us to go well beyond the point where large-scale
scaling is reached, and even to reach small-scale scaling (under certain 
circumstances).

We must emphasise that  we have chosen initial conditions for the radiation
and matter era boxes for which we knew (by performing other smaller simulations
beforehand \cite{moore,prl} and by using analytic
modelling \cite{ms1b,extend}) that the approach to scaling would
be fastest. Also with this aim, we have started the runs giving the strings
an initial RMS velocity $v_i\sim0.6$. For other choices of initial conditions,
the approach to scaling would be considerably slower. By the end of each
run, the error made in energy conservation was less than half of a percent.

\begin{figure}
\includegraphics[width=3.5in,keepaspectratio]{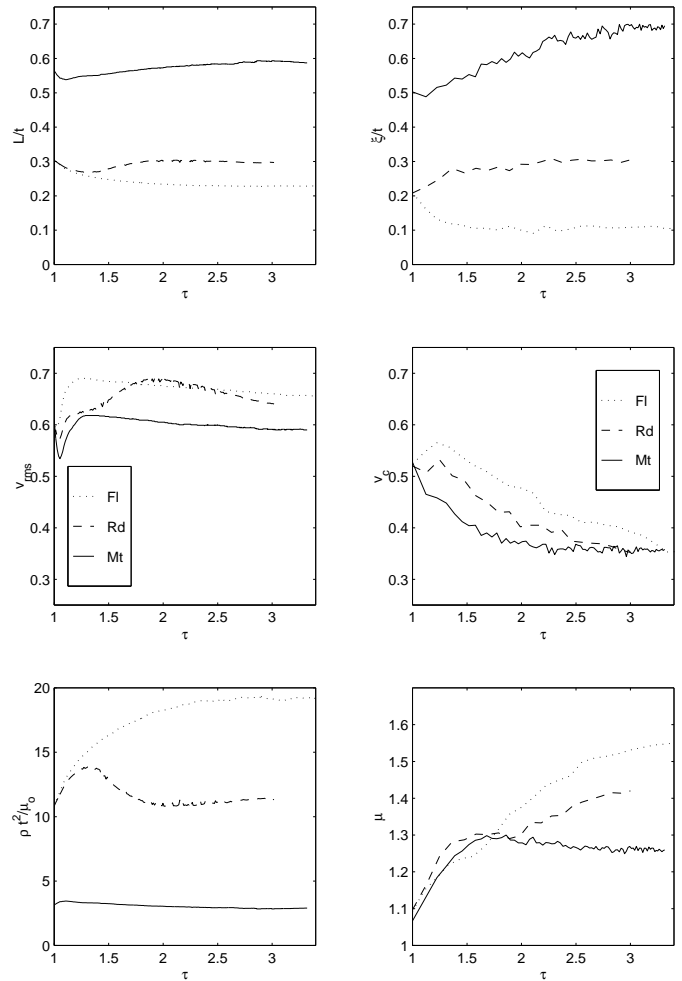}
\caption{\label{figlg}The time evolution of some characteristic
large-scale properties of cosmic string
networks, for 75 PPCL matter (solid lines), radiation
(dashed) and flat-spacetime (dotted) runs. In all plots the horizontal axis
represents the dynamic range in conformal time. All plotted quantities are defined in the text.}
\end{figure}

Consider first some averaged properties of the networks, displayed in
Fig.~\ref{figlg}. This shows the evolution of
the `characteristic' lengthscale
\be
\rho_\infty\equiv\mu_0/L^2\,,\label{bigl}
\ee 
with $\mu_0$ being the `bare' string mass per
unit length), the correlation (or persistence) length $\xi$ (to be defined below), the RMS string
velocity, the coherent velocity at the scale of the correlation
length $v_c$, the dimensionless parameter $\rho_\infty t^2/\mu_0$
(essentially a measure of the number of long strings inside one horizon
volume), and the `renormalised' dimensionless string mass per unit length
defined as
\be
\mu_{\rm{eff}}\equiv \mu \mu_0\,\label{mueff}
\ee
at the scale of the correlation
length. Note that the fact that we have started
the runs with a non-zero string velocity is responsible for giving
the initial network an effective mass per unit length different from unity.
Note also that while $L$ is simply just a measure of the string energy in a given volume, $\xi$ is the actual correlation length, and it is measured directly from the simulation. 
It is obvious these properties have strikingly different
behaviours in the three different regimes.
Interestingly, the only property which seems to be relatively independent
of the cosmological scenario is the coherent velocity at the scale of the
correlation length.

\begin{figure}
\includegraphics[width=3.5in,keepaspectratio]{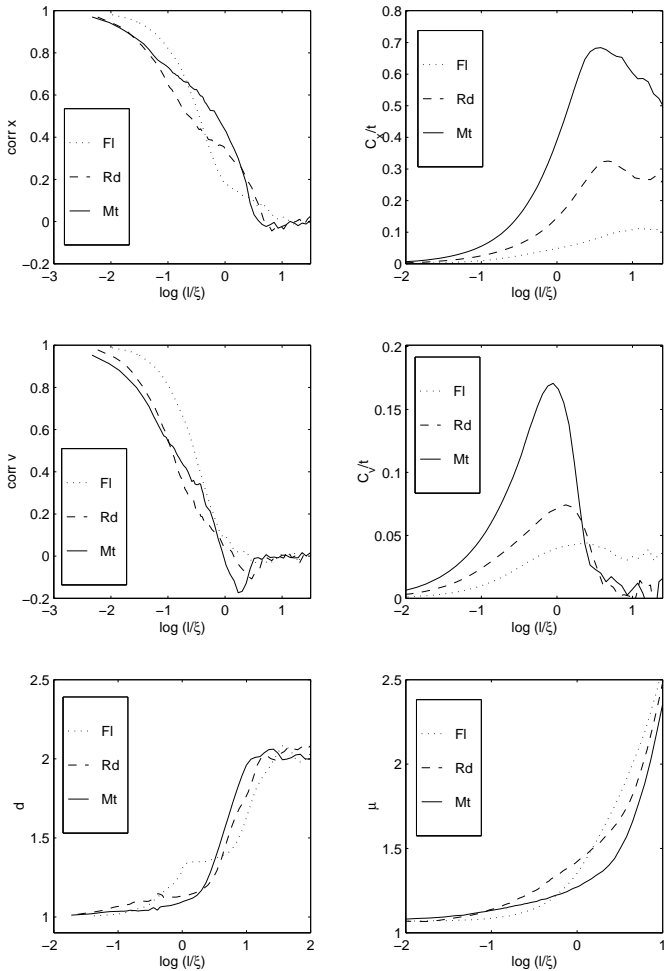}
\caption{\label{figfrac}Some characteristic small-scale properties of
cosmic string
networks in the linear scaling regime, for 75 PPCL matter (solid lines), radiation
(dashed) and flat-spacetime (dotted) runs. In all cases, the properties are
measured after evolution by a dynamic range of about 3.
In all plots the horizontal axis
represents the logarithm of the physical lengthscale relative to the
correlation length of the network. All plotted quantities are defined in the
text.}
\end{figure}

We have also performed a detailed analysis of the
\textit{small-scale} properties of these networks. This involves a
significant amount of data analysis, the details of which will be provided
elsewhere \cite{prep2}. Here, we will simply analyse a typical
`late-time' box of each of the three simulations above. 
(The analysis of the time evolution
of these properties for different epochs in the approach to scaling can
also be done.) In order to have a more
meaningful comparison, we have chosen boxes that have evolved
for a dynamic range (in conformal time) just above $3$ (the latest
time-step we obtained for the radiation era run).
We have exhaustively analysed boxes for other nearby time-steps,
and find that the results to be presented below remain unchanged,
within errors, so we can confidently say that these properties are indeed
representative of the epoch in question (with one exception that we will
point out below). The results are
presented in Fig~ \ref{figfrac} as a number of quantities measured on a
range of physical scales \textit{relative to the correlation length $\xi$}.
Each point on the graphs results from an average over (typically) a few
thousand measurements performed on randomly selected points on the simulation
box. The plots do not show the entire range of scales we probe---only those
which we believe to be accurately sampled and free of systematic errors
(e.g., box-size effects).

The top two panels show the correlation function for the tangent vectors,
\be
corr_x(s)=\langle{\bf x'}(s_0)\cdot {\bf x'}(s_0+s)\rangle\, \label{defcorr}
\ee
and its cumulative sum ($C_x(s)=\int_0^s corr_x(s')ds'$, plotted relative to
physical time $t$). Notice the differences between
the three cases for scales around and sightly below the correlation length.
In particular, this confirms that the correlation length $\xi$ (which is defined as the smallest scale at which the correlator vanishes, and is therefore roughly the height of the peak in the $C_x/t$ curve) is quite different
in the three cases, as we already saw in Fig.~\ref{figlg}. The middle panels
show the analogous functions for the velocity vectors,
${\bf {\dot x}}$. A first striking feature is that in the
expanding universe cosmic string velocities
are \textit{anti-correlated} on scales between
the correlation length and the horizon. However, such a feature is not present
in flat spacetime. This anti-correlation is the result of a `memory' of the
network for recent intercommutings, and its absence in the flat spacetime
case highlights the fact that the loop production mechanism is different
in the expanding and non-expanding cases. Indeed, such an effect was discussed
in \cite{ack}  (see also \cite{asconf}). If one defines a `velocity coherence length', this will be significantly smaller than $\xi$ itself.

Finally, the bottom two panels show what we believe to be
the most fundamental microscopic properties
of the string networks. On the left we have plotted the generalized fractal
(also referred to as `multi-fractal' or `differential fractal', depending on the context) dimension of the network \cite{frac1,frac2,frac3},
again as a function of scale relative to the correlation length.
On the right we have the `renormalized' (dimensionless) string mass per unit
length, $\mu$. The fractal dimension is obviously
unity on small scales (strings are roughly straight lines) and two on
very large scales (strings are roughly random walks). The interesting
question, however, is what happens at intermediate scales. In particular, one should expect a range of scales where strings should behave as \textit{self-avoiding random walks}, and these have a fractal dimension $d_s=5/3$ in three spatial dimensions.

We can immediately see that the radiation and matter era boxes have strikingly
similar fractal profiles (recall that we have rescaled by the respective correlation lengths, so the physical scales involved will be different). The differences result in a slightly higher mass per unit length on
a given scale (as one can confirm on the right). The flat spacetime profile,
however, is qualitatively different, and shows a rather high `step' around
the scale of the correlation length.   We have evidence that this step in the fractal
dimension tends to diminish and move
very slowly to smaller scales as the simulation progresses, indicating that 
flat space simulations have not really achieved small-scale structure 
scaling. Nevertheless, we
have performed moderately high resolution simulations with a dynamic range of 10 and the `step' is still clearly visible, so we believe it is to some extent a persistent 
feature.  This difference in the fractal properties must reflect the 
differing efficiencies of loop production in flat space and the expanding universe.  
While the integrated loop production efficiency is much greater in flat spacetime
$c=0.57$, as opposed to the expanding $c=0.23$ \cite{moore,prl}, it appears to be relatively 
less effective around the correlation scale with energy `stuck' on fairly large scales.  
This intuitive picture
can be confirmed by noting that the renormalised mass per unit length $\mu$
is smaller than in the expanding case on small scales, but is larger
on large scales.

Relevant to this issue, we note that a different (functional forms) approach to 
flat spacetime simulations was
recently presented by \cite{olum}.  Simple comparisons indicate that
their results appear to be in good agreement with our flat spacetime ones, though
note that an allowance has to be made for the fact that the two codes use
different definitions when calculating the correlation length $\xi$.  
Another difference is that our characterization of the statistical
properties of the network is based on fractal properties, whereas theirs
is based on Fourier space techniques. In fact, there are very interesting
relations between the two approaches, that will be explored elsewhere \cite{prep2}.

Finally, we point out that in all cases the fractal
dimension of the network at the scale of the correlation length is well
below two: typical values are 1.2 in the expanding case
and 1.4 in the flat case. (Note that this is to be expected if one interprets $\xi$ as a persistence length.)
So the intuitive picture that a string network looks Brownian at the scale
of the correlation length is clearly incorrect. Indeed, it's not even
strictly true at the scale of the horizon---here the network looks more like a self-avoiding random walk (an obvious consequence of intercommutings). The Brownian picture is only valid on significantly larger scales.
This is relevant in a number of contexts, in particular for the `toy model'
approach discussed in the introduction, which implicitly models the network
under the former assumption.

\section{\label{scal}When are we scaling?}

\begin{figure}
\includegraphics[width=3.5in,keepaspectratio]{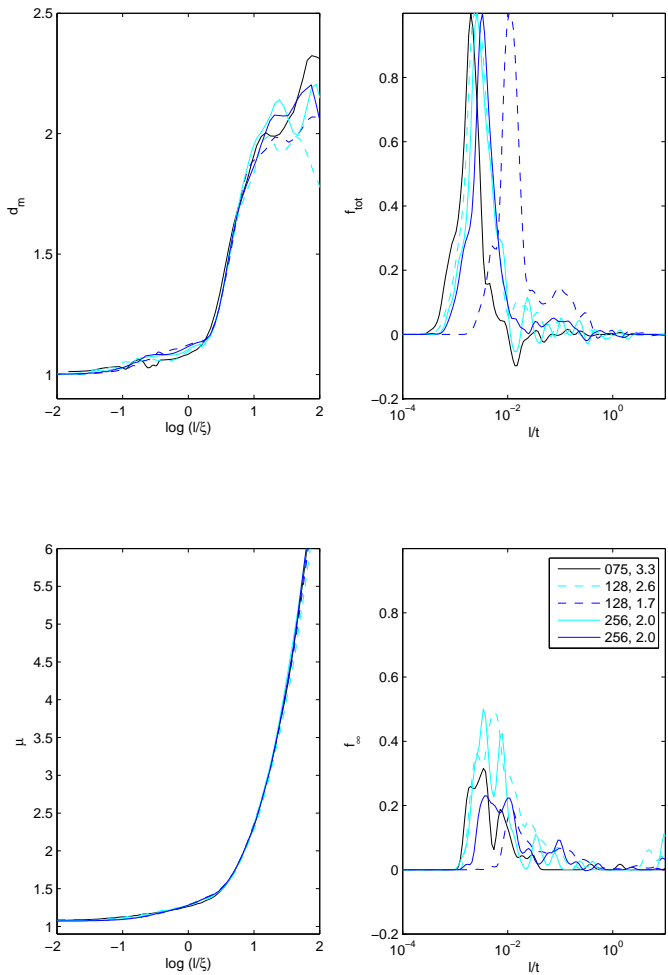}
\caption{\label{loopone}The multi-fractal dimension (top left), renormalized mass per unit length (bottom left), overall net loop production function (top right) and net loop production function from long strings only (bottom right) for a series of 5 different matter era runs. The resolutions (in PPCL) and dynamical ranges (in conformal time) are indicated in the legend.}
\end{figure}

The key question is now whether one can see scaling of the small-scale features of the cosmic string network. Detailed analysis indicates that quantities like the multifractal dimension  $d_m$, renormalized mass $\mu$ or coherent velocity as a function of scale evolve significantly at the beginning of the simulations, and even continue to do so after the large-scale properties of the network reach scaling (say, as measured by $L/t=const$, so that there in a constant number of strings---about 40---per Hubble volume). On the other hand, if one has high enough resolution (to probe lengthscales well below the correlation length) as well as dynamical range, one will see the evolution of $d_m(\ell)$ and $\mu(\ell)$ stop, hence achieving scaling. The first evidence for this is shown in the left panels of Fig.~\ref{loopone} where these quantities are plotted for the late-time evolution of five different matter era runs. Note that  the initial conditions in each are very different. Not only do the resolutions vary between 75 and 256 PPCL, but the 128 and 256 PPCL boxes  are two very different simulations, one with $v_i=0$ and the other with $v_i=0.6$ (and with different initial densities). Despite this, one sees that the $d_m(\ell)$ and $\mu(\ell)$  profiles are virtually identical, within errors. We have similar evidence of scaling for the radiation era, although the convergence is somewhat slower---this will be discussed in more detail elsewhere \cite{prep2}.  

These matter era results, in particular, now offer unambiguous evidence that small-scale 
structure has achieved scaling on all lengthscales accessible to these high resolution numerical simulations. As indicated by and conjectured from an earlier study \cite{asconf}, it now seems clear that 
the dynamical processes of reconnection and loop production are sufficient to govern the build-up of small-scale structure on cosmic string networks, without requiring additional damping mechanisms such as gravitational radiation.  Of course, simulations of even higher resolution and 
longer dynamical range will inevitably acquire small-scale structure on even smaller scales,
but it seems reasonable to suppose, given the convergence of the renormalised tension, that the 
integrated effect of these additional modes will be relatively small.  
We note that similar evidence for the scaling of small-scale structure on particular 
lengthscales has also been presented from flat spacetime simulations \cite{olum}, though 
this is somewhat weaker because, like our own flat space results, a wide range of other lengthscales continue to evolve slowly in this case.

\begin{figure}
\includegraphics[width=3.5in,keepaspectratio]{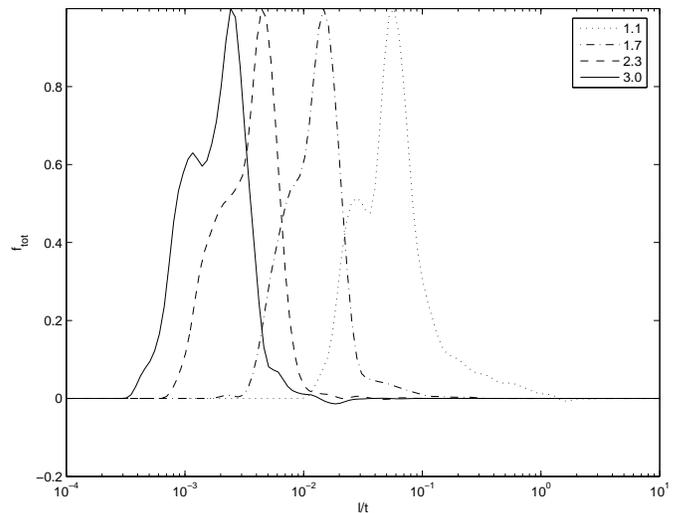}
\caption{\label{loopvar}The overall net loop production function for different times in the evolution of the 75 PPCL matter era run. The dynamical range of each plot, equally spaced in terms 
of horizon times, are indicated in the legend.}
\end{figure}

A related but different issue is that of loop production. On the right-hand panels of Fig.~\ref{loopone} we quantify the net loop production function of each of these networks for the timesteps corresponding to the left-hand plots. We show both the overall net production function (top panel, summing all loop creating or destroying events) and the production from long strings only (bottom panel).  The dominant overall loop production mechanism is that of the self-intersection of larger loops, so note that direct loop production by long strings typically happens on larger lengthscales (above the resolution of the simulations).  It is important to note that these are the time-dependent loop production 
functions, that is, the loops produced or destroyed on each lengthscale in a very narrow time span centered around the corresponding timestep.  This is a much finer diagnostic of loop generation mechanisms than the overall string loop density, and it also 
indicates the typical lengthscales of the long-lived loops which are relevant, for example, for gravitational 
wave background estimates.  Unfortunately,  
Fig.~\ref{loopone} indicates that the dominant loop production scale remains at least
 weakly dependent on the simulation resolution -- there is not the same strong agreement 
as between the fractal profiles on the left-hand side.  This is even clearer from the equally-spaced time evolution of the loop production function shown in Fig.~\ref{loopvar} for  the 75 PPCL matter era. This scale starts out being about the size of the correlation length, but becomes progressively smaller as small-scale structure builds up on the strings.  The evolution of the  peak of the loop distribution, however, 
is clearly beginning to slow down at late times indicating that it is rising above the minimum simulation resolution and will approach scaling.

It is clear from Fig.~\ref{loopone} that even though all the networks are statistically almost identical on a wide range of lengthscales, the dominant loop production scales are much more 
sensitive to variations on very small scales. Indeed we find that even when small-scale structure has reached scaling (as measured by $d_m(\ell)$, $\mu(\ell)$ or other diagnostics to be discussed in \cite{prep2}) the typical loop production scale can still evolve significantly towards smaller scales. However, we do have some evidence that this migration is significantly slowing down, particularly in the matter era, so that the loop production scale will itself reach scaling.

\section{\label{sdsc}Discussion and conclusions}

We have presented results from  a detailed analysis of the small-scale and loop production properties of cosmic string networks, in both flat spacetime and the expanding
universe. We have shown that, while the differences between the radiation and
matter epochs are largely `quantitative', the evolution in flat spacetime
is fundamentally different from the expanding case.  This indicates that flat spacetime
simulations, while qualitatively useful, will not provide an accurate approximation 
for cosmic strings in an expanding universe.

The two distinguishing characteristics of string evolution in Minkowski
spacetime are the absence of velocity anti-correlations on scales around
the correlation length, and the apparent existence of a `preferred' scale (roughly
around the correlation length $\xi$) from which
energy does not move to smaller scales.
We believe that these can have a substantial influence, for example when
calculating unequal time correlators for the network, which we are currently 
investigating in more detail.   Another result of our detailed analysis
is that cosmic string networks are only random walks on length scales
well outside the horizon, indicating that the semi-analytic models for strings 
need to be improved.

Finally, from high resolution matter era simulations we have presented comprehensive 
evidence over all lengthscales that small-scale structure converges to a scaling solution, that is, through reconnections alone without the need for additional damping mechanisms such as gravitational 
radiation.  Evidence for similar convergence has also been found from radiation era 
simulations.   While loop production is clearly more sensitive to the presence 
of structure on the smallest scales probed by the simulations, there are good indications that this will also achieve scaling.  We will present a more detailed and quantified analysis of these results elsewhere \cite{prep2}.    

\begin{acknowledgments}
We would like to thank Pedro Avelino, Brandon Carter, Tom Kibble, Ken Olum and Ladislav Skrbek for useful discussions. This work was performed in the context of
the ESF COSLAB network, and funded by FCT (Portugal), through grant POCTI/CTE-AST/60808/2004, in the framework of the POCI2010 program, supported by FEDER.   This 
work was also funded by PPARC grant PP/C501676/1.
The string code employed was developed by EPS in collaboration with Bruce Allen 
\cite{as}.   The numerical simulations were performed on the COSMOS,
the Altix3700 owned by the UK
Computational Cosmology Consortium, supported by SGI, Intel, HEFCE and PPARC.
\end{acknowledgments}

\bigskip
\noindent
{\bf Note added:} While our paper was being completed, ref.~\cite{Ringeval} appeared on the archive.
This paper presents some evidence for the scaling of the overall loop distribution on 
intermediate length scales below the correlation length but still near it, roughly loop lengths
 $\xi/10$ -- $\xi$.  A comparison with the net loop production function (Fig.~\ref{loopvar}) 
indicates that a population of short-lived relatively large loops is being characterised.  These 
loops will soon cascade through self-intersection down to much smaller sizes, presumably 
determined at present by their simulation resolution.

\bibliography{fractal}
\end{document}